\documentclass[a4paper]{jpconf} 
\usepackage{cite}
\usepackage{graphicx}
\usepackage{float}
\usepackage{url}
\usepackage{amsmath}

\begin{document}

\title{Development of a computational software in Python, used to study the materials resistance in beams}
 
\author{Julian A Rivera S$^1$ and Alex F Estupi\~n\'an L$^{1}$}
   
\address{$^1$ Universidad de Investigaci\'on y Desarrollo, Bucaramanga, Colombia.} 

\ead{aestupinan4@udi.edu.co}
 
\begin{abstract}
In this research, we do a software writing in Python to calculate the efforts, bending moments and deformations in beams of different materials. This computational tool, that we developed, is of great help in area of computational physical, more exactly in resistance of materials, which serves as support for researchers and teachers especially in physics and civil engineering, of the part of statics, who wish to carry out the modeling of the functions, involved in the calculation of resistance in beams in a practical and simple way, using the software presented in this article. In order to carry out this software, we are going to use the following methods: the double-integration method and the conjugate-beam method, which will serve as the basis of calculation to find the mathematical expressions involved in the analysis of resistance in beams, then we will perform the implementation of the aforementioned methods, using Python as the programming language. As a final step in this project, the graphical interface of said calculation tool will be made using the Python 3.0 Tkinter library. In this work, we show the results of the graphs of the stress profiles, bending moments and deformations, for the case of different types of beams, load and force distributions applied to them. Where we were also able to conclude that a calculation software was successfully built, dedicated to the analysis of efforts and deformations in beams made of different materials.
\end{abstract}
 

\section{Introduction}

The study of efforts and deformations in beams, mainly in the area of materials resistance, has seen a great growth in research in the last twenty years, due to the search for new elements and components in construction, which can offer greater reliability and security to large construction companies around the world \cite{cite_1,cite_2,cite_3,cite_4,cite_5}. 

It's for this reason, the importance of the study of the deformations suffered by the beams should be highlighted, in addition to being able to analyze how this wear depends mainly on the type of material that was used in the construction of the beams, which has led researchers to seek new materials and more resistant and durable alloys. One of these efforts by researchers has focused on understanding the behavior of Fiber-Reinforced Polymer (FRP) reinforced concrete structures \cite{cite_6,cite_7}. These types of materials, commonly referred to in the industry as fine materials, have presented significant advantages over traditionally used construction materials, such as: steel, wood and concrete \cite{cite_8,cite_9}.

As a consequence of the appearance of these new materials, along with this, the concern of some researchers for rheological studies has been presented, such as the evaluation of the resistance, durability and reliability of these new and innovative materials in the construction, it is necessary to evaluate both theoretical, experimental and computational, stress tests, deformation and presence of bending moments by loads on the beams of these materials \cite{cite_10,cite_11,cite_12}.

At present, several investigations have been presented about a great effort, in the development of the calculation of stresses, stresses and moments of stress in beams, one of the main pioneers were, Trantino and Dezi \cite{cite_13}, the culaes pose in his work a time-dependent analysis of composite steel and concrete beams with flexible shear connectors \cite{cite_14}.

On the other hand, more recent studies \cite{cite_15,cite_16,cite_17}, in which the numerical models proposed by (Amadio et al. 1999) have been taken as a starting point. These models have shown a very good estimate of the damage and possible wear that long-term beams may suffer, especially those made of concrete, cement and wood \cite{cite_18}.

In this work we develop a computational code, which is capable of reproducing the calculation of the stresses, bending moments; with the objective of being able to predict the possible zones or places of failure, which the beams may present that are being evaluated.

\section{Methodology}

With the purpose of developing this software for calculating efforts and deformations in beams, the authors have followed the following steps to carry out this study, which were:

\begin{enumerate}
	
	\item Review of the theoretical framework of calculation methods.
	
	\item Implementation of calculation methods in a computational code written in Python.
	
	\item Construction of the graphical interface, to facilitate access and improve the presentation of the computational calculation tool developed.

\end{enumerate}

The first thing that we must take into account, is the type of structure that we are studying, we know that there are mainly two types, which are hyperstatic and isostatic \cite{cite_19}. In the type of hyperstatic structures, there are a greater number of forces that act on the beam than equations that are in equilibrium, which immediately presents us with a problem, in order to find the efforts that act on it; additional equations need, with the displacements or turns at a specific point to know these forces and in this way be able to raised the equations known as: compatibility equations \cite{cite_20}.

On the other hand, the type of isostatic structures are those that their reactions can be calculated using the static equations for the forces and bending moments, that is, in a more technical way we can say that an isostatic structure has the same number of equations as unknowns, therefore, it can be solved using a system of linear equations \cite{cite_21,cite_22}.

In our research, we want to analyze and study the behavior of the hyperstatic beams also known as elastically indeterminate, in this case we will need the deformations that a beam undergoes when placing a certain amount of load. Where we know that these loads will generate bending and shear stresses in the beam, which will cause the beam to undergo certain deformations.

In this way, our two most relevant objectives in this study are: first, to obtain new conditions for analysis, which will offer us more equations to be able to solve our system of equations of more unknowns than equations, as a second objective will be to be able to evaluate and design the beams, according to the requirements under the load conditions to which the beams will be subjected, always taking into account the conditions requested by design and construction that is to be carried out.

Our purpose and impact in the area of physics, of our research project, is based on being able to contribute and offer a calculation tool that is accessible to researchers who are concerned and dedicated to the study and analysis of the resistance of materials in recessed and simply supported beams.

\section{Theoretical framework}

The two methods most used to do the analysis of deformations in beams, and wich we use in this work, are:

\begin{enumerate}
	\item Double-integration method.
	\item Conjugate-beam method.
\end{enumerate}

In each of the methods listed above, we are concerned with determining the angle of curvature of the elastic line and its respective deflections or arrows. It is very important to highlight that depending on the type of beam that is being studied, it is more convenient to use any of the analysis methods previously mentioned.

\textbf{(i) Double-integration method:} This method mainly consists of finding the expressions of the slope functions $\theta(x)$, the bending moment $M(x)$ and the deflection $y(x)$, starting from Equation (\ref{eq_9}), also known as Differential equation for the elastic curve of a beam \cite{cite_25,cite_26}.

\begin{equation}
EI \frac{d^2y(x)}{dt^2} = M(x),
\label{eq_9}
\end{equation}

We must take into account; that by solving equation (\ref{eq_9}), we are going to obtain an expression of the deflection of the beam $y(x)$, as a function of two arbitrary constants, which will be found using the boundary conditions. Keep in mind the importance of calculating the deflection of the beam, which consists mainly of being able to find the maximum deflection and determine the value of the reactions of the beam.

We can also relate, the expression of the load $W(x)$ and the shear $V(x)$, as a function of the bending moment $M(x)$, from equations (\ref{eq_10})-(\ref{eq_11}), respectively \cite{cite_25,cite_26}.

\begin{equation}
W(x) = \frac{d^2M(x)}{dx^2},
\label{eq_10}
\end{equation}

\vspace{-0.1cm}

\begin{equation}
V(x) = \frac{dM(x)}{dx}.
\label{eq_11}
\end{equation}

In some cases and problems of the beam analysis study, more than two boundary conditions are needed, to calculate the unknown reactions that are applied to the beam, due to this reason, we can notice that we need redundant relationships, generally the number of Boundary conditions, necessary to fully return, a study of stresses and deflections in the beams, is determined by Equation (\ref{eq_12}).

\begin{equation}
B.C = n + 2,
\label{eq_12}
\end{equation}

where $B.C$ is the number of boundary conditions and $n$ is the degree of statically indeterminacy of the beam.

\textbf{(ii) Conjugate-beam method:} This method of analysis of stresses in the beams, consists mainly in solving the system of equations, shown in equations (\ref{eq_13})-(\ref{eq_14}), which allow us to have a relationship between the loading, shear, and bending moments\cite{cite_27,cite_28}.

\begin{equation}
\frac{d^2M(x)}{dx^2} = \frac{dV(x)}{dx} = w(x),
\label{eq_13}
\end{equation}

\vspace{-0.1cm}

\begin{equation}
\frac{d^2v(x)}{dx^2} = \frac{d\theta(x)}{dx} = \frac{M(x)}{EI}.
\label{eq_14}
\end{equation}

Where $M(x)$ is the bending moment; $V(x)$ is the shear; $w(x)$ is the distributed load, $\theta(x)$ is the slope and $V(x)$ displacement of the real beam. Thus, we can see that to use these equations, an analysis of an imaginary beam must be taken, which has the same dimensions (length) as the original beam, but the load at any point of the conjugated beam; equals the bending moment at that point divided by the expression $EI$. Where this analysis should always be carried out under static equilibrium conditions.

\section{Results}

\begin{figure}[htbp]
  	\centering
  	\begin{minipage}{0.48\linewidth}
  		\centering
  			\includegraphics[width=\linewidth]{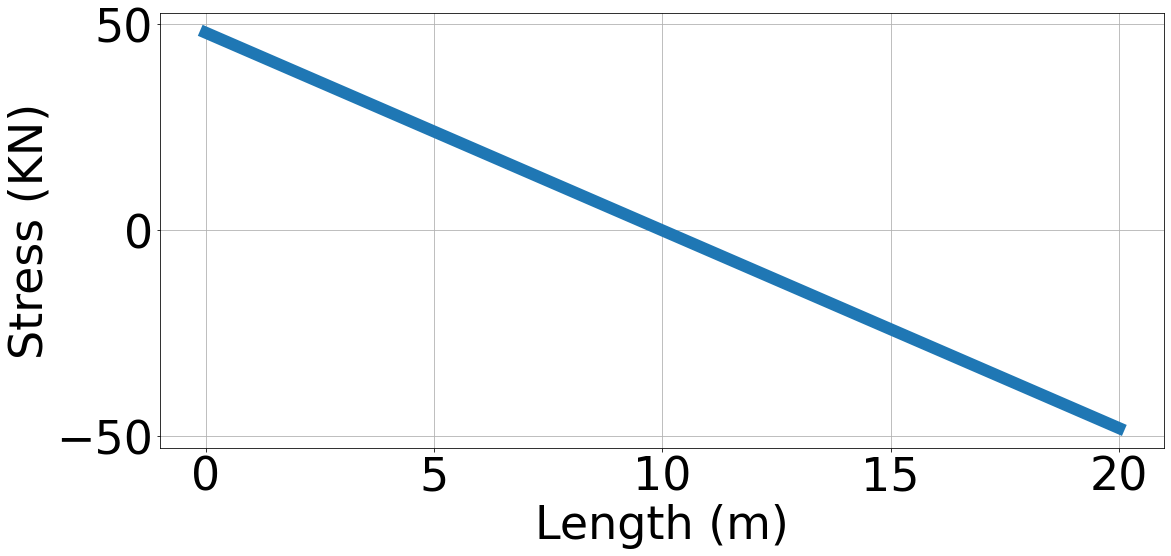}
			\\(a)
  	\end{minipage}
  	\hspace{0.01\linewidth}
  	\begin{minipage}{0.48\linewidth}
  			\centering
  			\includegraphics[width=\linewidth]{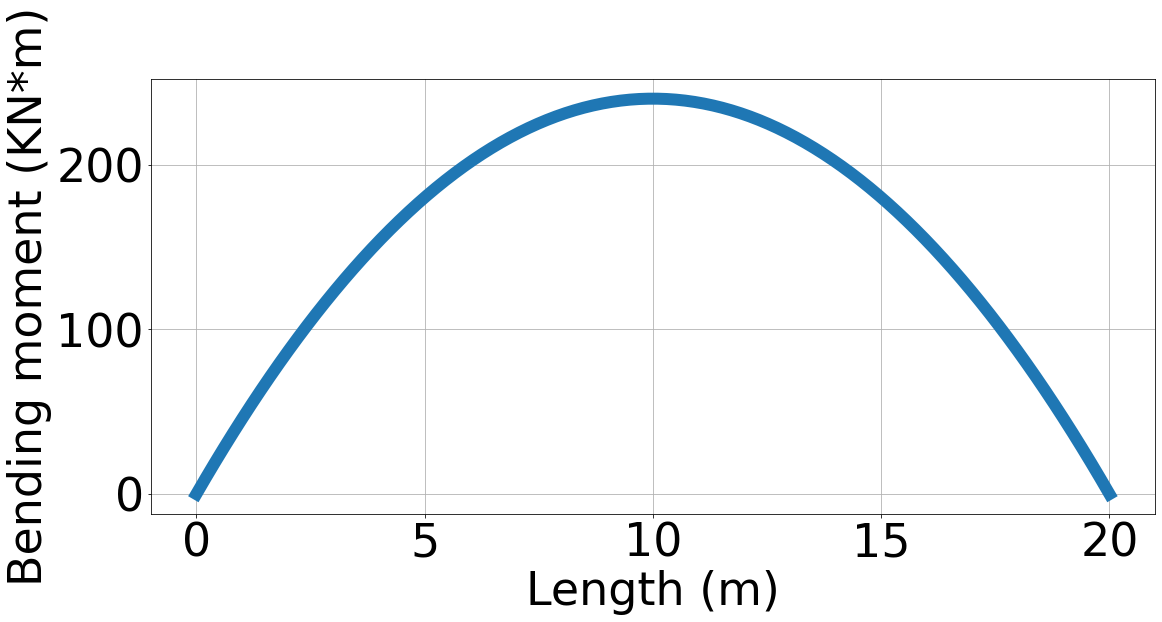}
  			\\(b)
  	\end{minipage}
  		\hspace{0.01\linewidth}
  		\begin{minipage}{0.52\linewidth}
  				\centering
  				\includegraphics[width=\linewidth]{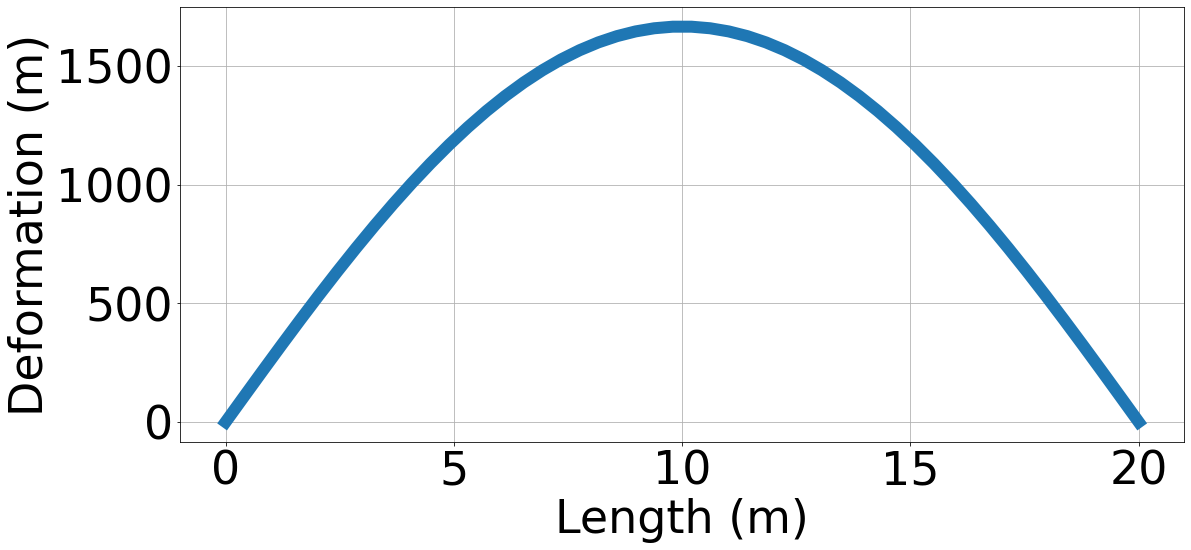}
  				\\(c)
  		\end{minipage}
   \caption{Analysis for a simply supported beam with a rectangular load distribution. (a) Stress as a function of length. (b) Bending moment as a function of length. (c) Deformation as a function of length.}
   \label{fig_1}
\end{figure}

In this section, we want to show the results that were obtained in the code written in Python 3.0, which will be used as a support tool, for researching in the area of materials resistance in physics. In this computational tool, we allow the possibility of calculating the effort as a first step, depending on the characteristics of the beam (beam dimensions, moment of inertia, elasticity index, etc.), in addition to the amount , shape and location in which the load is placed on the beam. Figure \ref{fig_1}(a), shows the behavior of the stress graph along the length of the beam, in the case of a beam simply supported with rectangular distributed load.

In this computational code that was written in this work, the study of the behavior of bending moments along the length of the beam was also carried out, in the case of a beam simply supported with rectangular distributed load, the bending moments in function of length are shown in Figure \ref{fig_1}(b).

Continuing, with the study of the analysis of stresses in beams, we want to show also the behavior of the deformation along the length of the beam, this behavior will depend largely on the value of the elasticity index of the beam material, which will depend mainly on the modulus of elasticity and the quality factor, these values will vary depending on the material to be used for the construction of the beam, which allows us to make a more realistic calculation of the expected values, from the deformation of the beam to along its length as shown in Figure \ref{fig_1}(c).

\begin{figure}[htbp]
	\centering
	\begin{minipage}{0.48\linewidth}
		\centering
		\includegraphics[width=\linewidth]{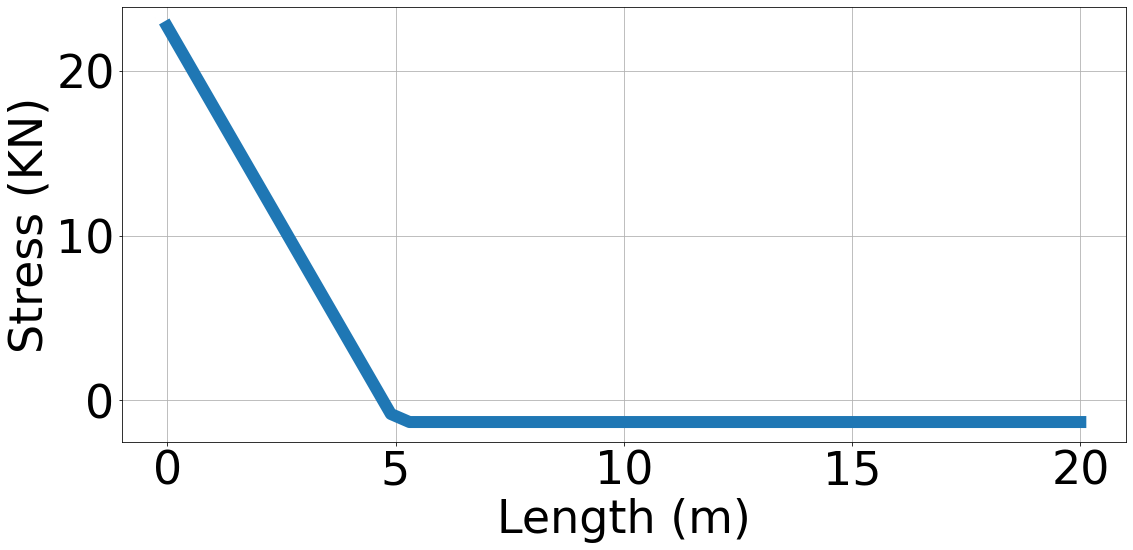}
		\\(a)
	\end{minipage}
	\hspace{0.01\linewidth}
	\begin{minipage}{0.48\linewidth}
		\centering
		\includegraphics[width=\linewidth]{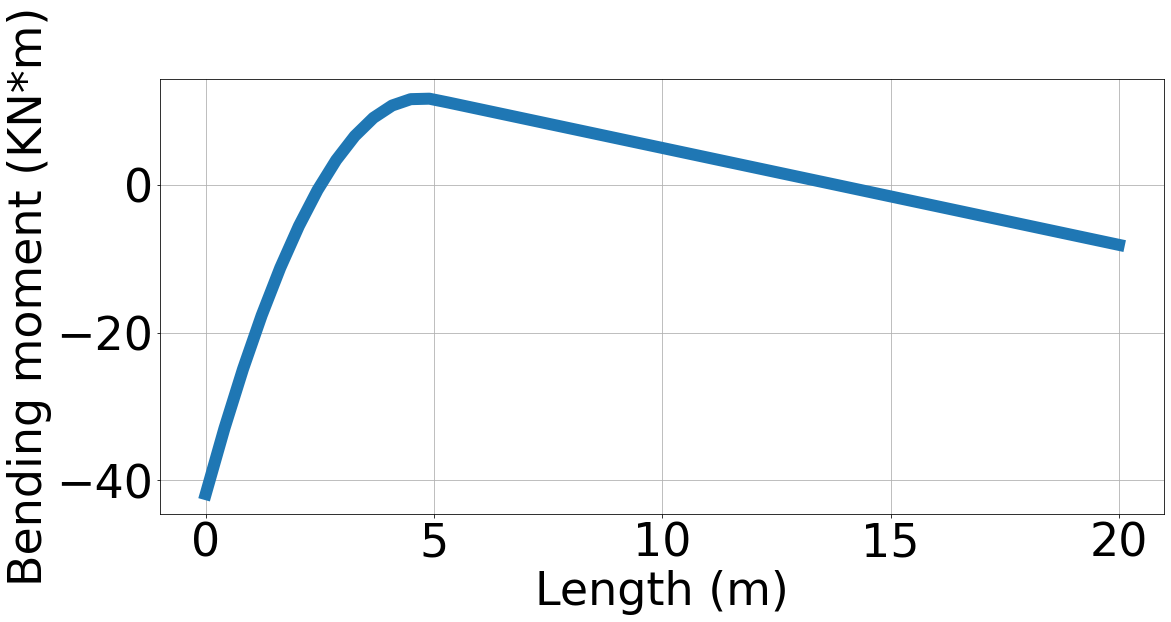}
		\\(b)
	\end{minipage}
	\hspace{0.01\linewidth}
	\begin{minipage}{0.52\linewidth}
		\centering
		\includegraphics[width=\linewidth]{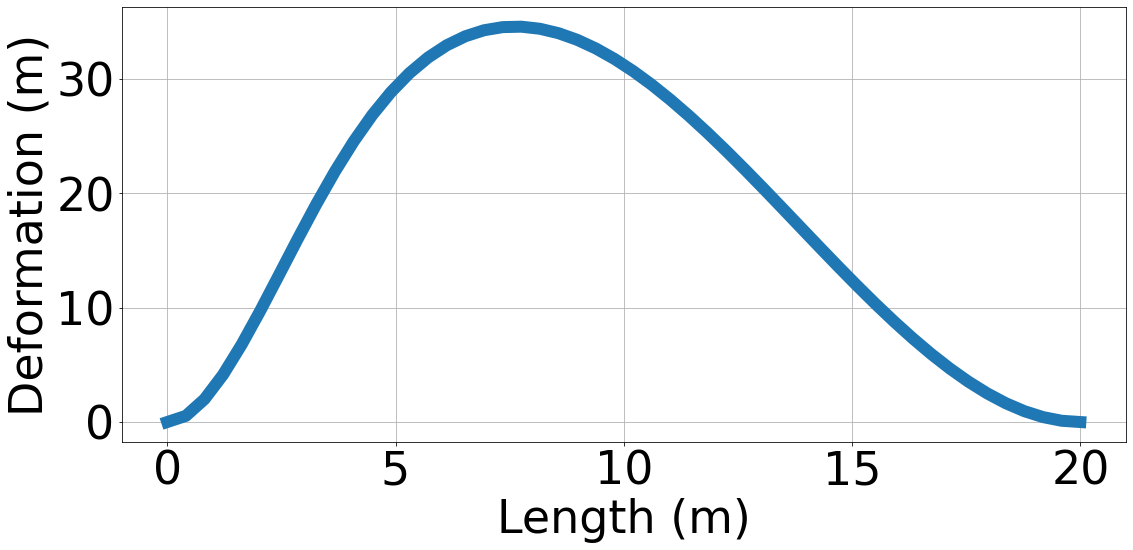}
		\\(c)
	\end{minipage}
	\caption{Analysis for a  simple embedded beam with a uniform load $q$ at the left end of it. (a) Stress as a function of length. (b) Bending moment as a function of length. (c) Deformation as a function of length.}
	\label{fig_2}
\end{figure}

On the other hand, in this research work the study of embedded beams was also carried out, in which the calculation of the stresses, bending moments and deformations for different configurations of embedded beams was performed. In Figure \ref{fig_2}(a), the behavior of the stresses along a simple embedded beam with uniform load $q$ at one end is shown, this graph shows a linear decrease in the stress on the beam, as it moves away from the point of start where the uniform load $q$ is placed on the beam (See Figure \ref{fig_2}(b)). To carry out the bending moment analysis, in a simple embedded beam with a uniform $q$ load at one end of it,  a maximum value of $11.5625$ $KN\cdot m$ was obtained, for a beam whose length is $20$ $m$, where the length in which the load is applied is $5$ $m$ and the value of the load applied on it is $4.8$ $kN/m$ (See Figure \ref{fig_2}(b)).

On the other hand, it is very interesting to see in Figure \ref{fig_2}(b), as in the section of the beam where the uniform load $q$ is not applied, the behavior of the bending moment of is linearly decreasing. Another very important analysis, which must be taken into account in the study of the resistance of materials, especially for the construction of a beam, is the study of deformations, along the length of the beam, for this reason. The Figure \ref{fig_2}(c), shows the deformation behavior of a simple embedded beam with a uniform load $q$ at the left end of it.

It is of great importance to highlight that Figure \ref{fig_2}(c), which shows the behavior of the deformation as a function of the length of the studied beam, mainly depends largely on the value of the moment of inertia $I$; and the elasticity index $E$ of the material with which you are working, the first parameter will depend on the geometric shape of the beam and the second parameter will depend mainly on the material in which the beam is built (for example: concrete, steel, iron, etc.). In the case of Figure \ref{fig_2}(c), a factor $E \cdot I = 6$ $kg \cdot m^2$ was taken, which corresponds to the product of the moment of inertia $I$ by the elasticity index $E$, in this specific case a maximum deformation value equal to at $34.596$ $m$ as shown in Figure \ref{fig_2}(c).

\begin{figure}[htbp]
	\centering
	\begin{minipage}{0.48\linewidth}
		\centering
		\includegraphics[width=\linewidth]{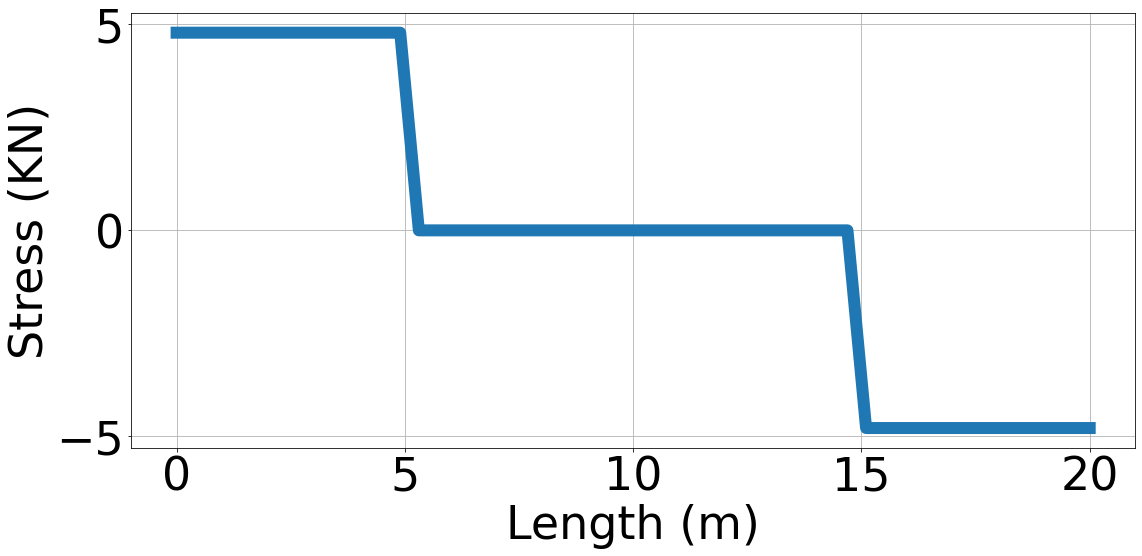}
		\\(a)
	\end{minipage}
	\hspace{0.01\linewidth}
	\begin{minipage}{0.48\linewidth}
		\centering
		\includegraphics[width=\linewidth]{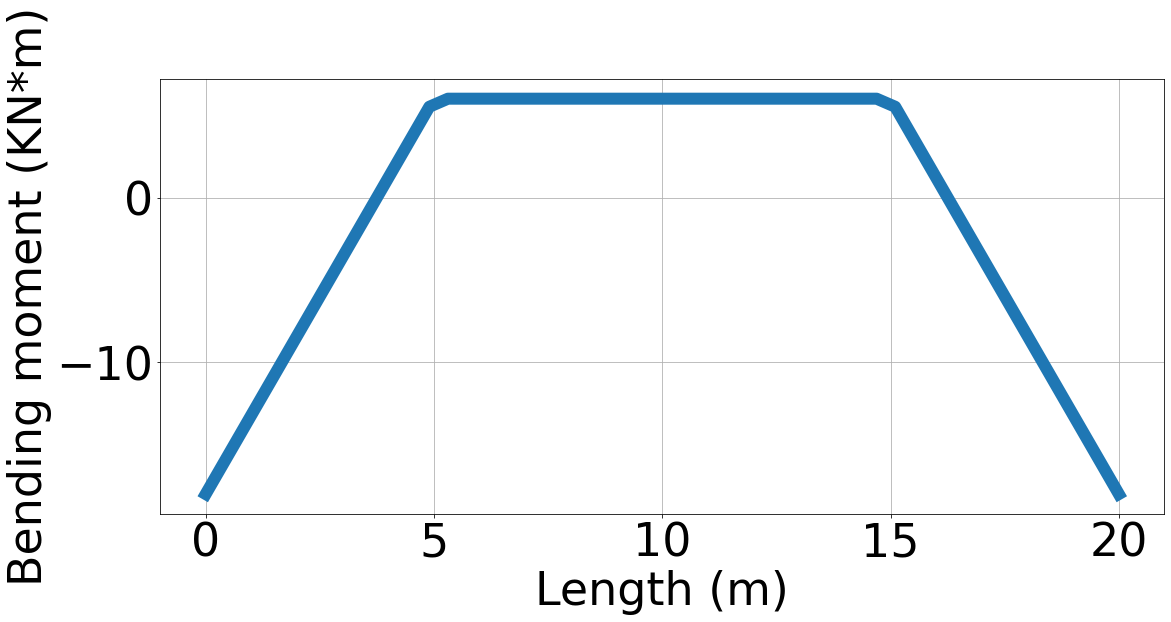}
		\\(b)
	\end{minipage}
	\hspace{0.01\linewidth}
	\begin{minipage}{0.52\linewidth}
		\centering
		\includegraphics[width=\linewidth]{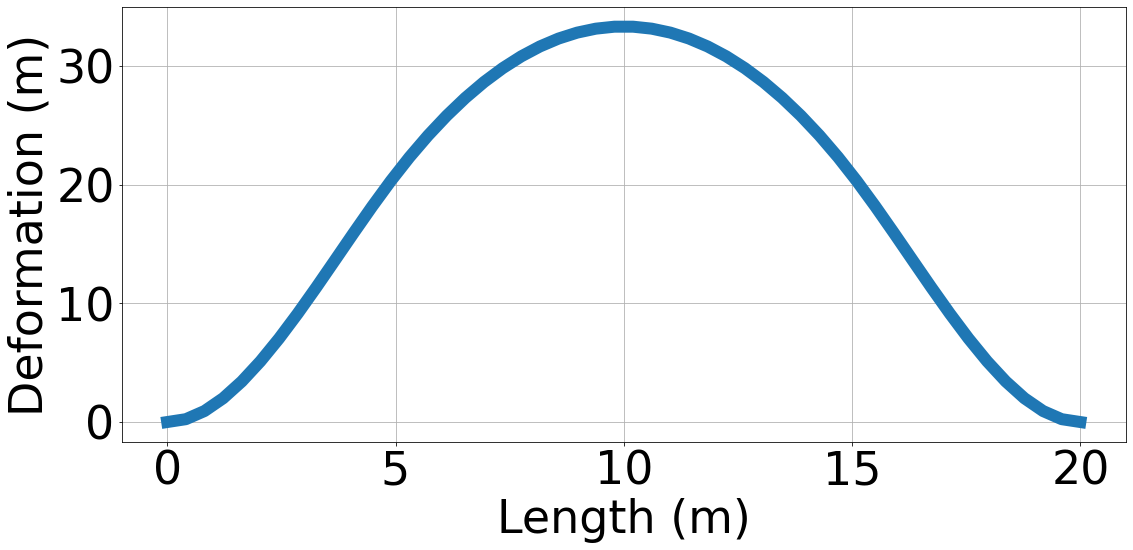}
		\\(c)
	\end{minipage}
	\caption{Analysis for a simple embedded beam, with symmetrical punctual $F$ loads applied on the beam. (a) Stress as a function of length. (b) Bending moment as a function of length. (c) Deformation as a function of length.}
	\label{fig_3}
\end{figure}

We also wanted to implement in our computational code, the stress analysis for the case of a single embedded beam, with symmetric point $F$ loads applied on the beam to be studied. Where it was obtained, that in the case of a beam, whose length is equal to $20$ $m$, in which two point loads of value $F$ $=$ $4.8$ $kN$ are applied, each $5$ $m$ from the ends of this beam, the values of the magnitude of the maximum stresses on this beam were $4.8$ $kN$, as seen in Figure \ref{fig_3}(a).

In Figure \ref{fig_3}(b), the behavior of the bending moment along the beam is shown, in which it can be seen that the maximum value of the bending moment on the beam is 6 kN * m, where this value remains constant mainly in the center of the beam (See Figure \ref{fig_3}(b)).

Finally, in Figure \ref{fig_3}(c), the behavior of the beam deformation is shown, as a function of length, where a peak or maximum value equal to $33.3$ $m$ can be seen in this graph, which is located right in the center of the beam, in this case in which the length of the beam is $20$ $m$, the peak is located $10$ $m$ from the origin of the beam as shown in Figure \ref{fig_3}(c).

\section{Conclusions}

In this work, it was possible to develop a virtual and computational tool written in Python 3.0, with which the stresses, bending moments and deformations of different beam configurations can be calculated for different geometric shapes of load and forces applied to them. Finally, it was possible to write a software, which allows us to perform an analysis of stresses and deformations in beams of different materials.

\section*{Acknowledgments}

The authors would like to express their thanks, especially to the Universidad de Investigaci\'on y Desarrollo \textit{UDI}, for all the human, material and financial support to carry out this research work.

\newpage

\section*{References} 

\bibliography{bibli.bib}

\providecommand{\newblock}{}
\begin{thebibliography}{10}
\expandafter\ifx\csname url\endcsname\relax
  \def\url#1{{\tt #1}}\fi
\expandafter\ifx\csname urlprefix\endcsname\relax\def\urlprefix{URL }\fi
\providecommand{\eprint}[2][]{\url{#2}}

\bibitem{cite_1}
Abeyaratne R and Knowles J~K 2000 {\em Journal of applied physics\/} {\bf 87}
  1123--1134

\bibitem{cite_2}
Chen Y~C and Lagoudas D~C 2000 {\em Journal of the Mechanics and Physics of
  Solids\/} {\bf 48} 275--300

\bibitem{cite_3}
Feng Z and Li D 1996 {\em Journal of Intelligent Material Systems and
  Structures\/} {\bf 7} 399--410

\bibitem{cite_4}
Cao W, Cudney H~H and Waser R 1999 {\em Proceedings of the National Academy of
  Sciences\/} {\bf 96} 8330--8331

\bibitem{cite_5}
KOISSTINEN D 1959 {\em Acta metallurgica\/} {\bf 7} 59--60

\bibitem{cite_6}
T{\"a}ljsten B 1997 {\em Journal of materials in civil engineering\/} {\bf 9}
  206--212

\bibitem{cite_7}
Triantafillou T~C 1998 {\em ACI structural journal\/} {\bf 95} 107--115

\bibitem{cite_8}
Ghobarah A 2001 {\em Proc., FRP Composites in Civil Engineering, CICE2001\/}
  {\bf 1} 701--712

\bibitem{cite_9}
Larbi A~S, Contamine R and Hamelin P 2012 {\em Engineering Structures\/} {\bf
  45} 12--20

\bibitem{cite_10}
Collet M, Folt{\^e}te E and Lexcellent C 2001 {\em European Journal of
  Mechanics-A/Solids\/} {\bf 20} 615--630

\bibitem{cite_11}
Junior F~S and Venturini W~S 2007 {\em Advances in Engineering Software\/} {\bf
  38} 538--546

\bibitem{cite_12}
T{\"a}ljsten B 2003 {\em Construction and Building Materials\/} {\bf 17} 15--26

\bibitem{cite_13}
Marcello~Tarantino A and Dezi L 1992 {\em Journal of Structural Engineering\/}
  {\bf 118} 2063--2080

\bibitem{cite_14}
Dezi L and Tarantino A~M 1993 {\em Journal of Structural Engineering\/} {\bf
  119} 2095--2111

\bibitem{cite_15}
Serrano E and Gustafsson P~J 2007 {\em Materials and structures\/} {\bf 40}
  87--96

\bibitem{cite_16}
Ceccotti A, Fragiacomo M and Giordano S 2007 {\em Materials and structures\/}
  {\bf 40} 15--25

\bibitem{cite_17}
Jorge L, Sch{\"a}nzlin J, Lopes S, Cruz H and Kuhlmann U 2010 {\em Engineering
  structures\/} {\bf 32} 3966--3973

\bibitem{cite_18}
Thirumalaiselvi A, Anandavalli N, Rajasankar J and Iyer N~R 2016 {\em Steel
  Compos Struct\/} {\bf 20} 167--184

\bibitem{cite_19}
V{\'a}zquez A~C, Navarro A~E and L{\'o}pez A~G 2012 {\em Scientia et
  technica\/} {\bf 1} 32--37

\bibitem{cite_20}
Kodur V and Dwaikat M 2008 {\em Cement and Concrete Composites\/} {\bf 30}
  431--443

\bibitem{cite_21}
Bailey C 1999 {\em Journal of Constructional Steel Research\/} {\bf 50}
  235--257

\bibitem{cite_22}
Steeves C~A and Fleck N~A 2004 {\em Scripta materialia\/} {\bf 50} 1335--1339

\bibitem{cite_25}
Shi J, Xu X, Wang J and Li G 2010 {\em Nondestructive Testing and Evaluation\/}
  {\bf 25} 189--204

\bibitem{cite_26}
Sigurdardottir D~H, Stearns J and Glisic B 2017 {\em Smart Materials and
  Structures\/} {\bf 26} 075002

\bibitem{cite_27}
Agarwal D 2018 {\em international journal of engineering trends and
  technology\/} {\bf 59} 63--65

\bibitem{cite_28}
Zhang S, Liu B and He J 2019 {\em Measurement\/} {\bf 133} 208--213

\end{thebibliography}
\bibliographystyle{iopart-num}

\end{document}